\documentclass[preprint2]{aastex}

\shorttitle{The Enigmatic HH 255}
\shortauthors{Matt \& B\"ohm}

\slugcomment{Submitted to \pasp.}

\begin{document}

\title{The Enigmatic HH 255}

\author{Sean Matt\altaffilmark{1}}
\affil{Physics \& Astronomy Department, McMaster University, Hamilton ON, 
Canada L8S 4M1}
\email{matt@physics.mcmaster.ca}

\and

\author{Karl-Heinz B\"ohm}
\affil{Astronomy Department, University of Washington, Seattle WA, USA 98195}
\email{bohm@astro.washington.edu}

\altaffiltext{1}{CITA National Fellow}

\begin{abstract}

%Burnham's nebula (aka HH 255) is part of the circumstellar material
%south of T Tauri and seems to be a part of the complex Herbig-Haro
%(HH) outflow from that star forming region.  The emission line fluxes
%in HH 255 agree quite well with the characteristic shock-excited
%spectrum of an HH object, but the spatially resolved and detailed
%spectra reveal some properties that do not agree with the standard bow
%shock model.

To gain insight into the nature of the peculiar Herbig-Haro object HH
255 (also called Burnham's nebula), we use previously published
observations to derive information about the emission line fluxes as a
function of position within HH 255 and compare them with the
well-studied, and relatively well-behaved bow shock HH 1.  There are
some qualitative similarities in the H$\alpha$ and [O III] 5007 lines
in both objects.  However, in contrast to the expectation of the
standard bow shock model, the fluxes of the [O I] 6300, [S II] 6731,
and [N II] 6583 lines are essentially constant along the axis of the
flow, while the electron density decreases, over a large distance
within HH 255.

We also explore the possibility that HH 255 represents the emission
behind a standing or quasi-stationary shock.  The shock faces upwind,
and we suggest, using theoretical arguments, that it may be associated
with the collimation of the southern outflow from T Tauri.  Using a
simplified magnetohydrodynamic simulation to illustrate the basic
concept, we demonstrate that the existence of such a shock at the
north edge of HH 255 could indeed explain its unusual kinematic and
ionization properties.  Whether or not such a shock can explain the
detailed emission line stratification remains an open question.

\end{abstract}


\section{Introduction \label{sec_introduction}}

Outflows from young stellar objects (YSO's) are complex and dynamic,
always resulting in violent interactions with circumstellar material
and internal interactions within the flow.  The most conspicuous
optical manifestation of these hypersonic interactions are Herbig-Haro
(HH) objects, which cool primarily by emission from hydrogen
recombination and collisionally excited, forbidden emission lines from
metals.  Undeniably, the study of these outflows is an important part
of understanding the star formation process as a whole \citep[for a
review, see][]{reipurthbally01}.

Burnham's nebula, called HH 255 by \citet{reipurth99} and condensation
E by \citet[][hereafter BS94]{bohmsolf94}, is part of the
circumstellar material south of T Tauri and seems to be a part of the
complex HH outflow from that star forming region.  A number of studies
have demonstrated that the forbidden line fluxes in HH 255 agree quite
well with the characteristic shock-excited spectrum of an HH object
\citep*[][hereafter SBR88]{herbig50, herbig51, schwartz74, solf3ea88}.
However, the spatially resolved and detailed spectra of HH 255 reveal
some properties that are difficult to understand, in the context of a
typical HH bow shock \citep[BS94;][hereafter BS97 and
SB99]{bohmsolf97, solfbohm99}.

The first obvious problem is that, while the high ionization lines of
[O III] and [S III] in HH 255 are quite similar to those of HH 1 (a
representative HH object) and require a shock velocity of $> 90$ km
s$^{-1}$, the measured radial velocity of HH 255 is very nearly zero,
with respect to the photospheric velocity of T Tau (SBR88; BS94).  One
may try to explain this problem by arguing that HH 255 is seen nearly
side on or that it is a stationary feature, such as a shocked ambient
cloudlet.  However, SBR88 and SB99 find an inclination of the
blue-shifted, southern T Tau outflow of $\sim 14^\circ$ and $11^\circ$
(respectively) from the plane of the sky, too large to account for the
small radial velocity.  The shocked cloudlet explanation (which
implies the existence of a bow shock facing toward T Tau) is ruled out
by the velocity dispersion observed in HH 255.  That is, the total
velocity dispersion (full width at zero intensity) of optical
forbidden lines in HH 255 is unusually small, less than 45 km s$^{-1}$
(BS97).  It is well known that bow shocks exhibit a large velocity
dispersion, comparable to (and most often greater than) the shock
velocity \citep*[see][]{hartigan3ea87, ragabohm86, choe3ea85}, as
material that enters the front of the shock is then accelerated
roughly perpendicular to the original flow direction and eventually
moves around the obstacle (e.g., the bullet, jet, or cloudlet).  In
addition, the velocity dispersion of the forbidden ionic and atomic
lines in HH 255 is the same as for molecular H$_2$ lines (SB99).  This
is surprising, since the H$_2$ lines form at much lower shock
velocity, and therefore typically come from a region with a smaller
velocity dispersion than the optical lines in HH objects.  Another
partial explanation would be that HH 255 results from a plane shock.
This could explain the low centroid velocity and velocity dispersion,
but BS97 pointed out that the cooling region would have a thickness of
only about 0\farcs1, while HH 255 shows nearly constant ionization
from about 4\arcsec--10\arcsec\ south of T Tau.  Also, the possible
existence of a plane shock in an HH flow lacks a theoretical
explanation.

The enigmatic ionization structure and kinematic properties of HH 255
prompt us to look in more detail at data obtained earlier from SBR88,
BS94, BS97, and SB99.  We do this to gain a more complete picture of
all observable properties to guide the models of this object.  In
addition, we further explore the possibility that HH 255 is a
stationary shock (that is neither a plane shock nor a bow shock).
First we review the known properties of HH 255 in section
\ref{sec_properties}.  In section \ref{sec_further}, we look at the
emission line fluxes as a function of position along HH 255 and
compare them to those of HH 1.  Finally, section \ref{sec_discussion}
discusses the interpretation of HH 255 as a standing shock, possibly
associated with the collimation of the outflow south of T Tauri.

\section{Known Properties of HH 255 \label{sec_properties}}

T Tauri is a binary system \citep*{dyck3ea82} at a distance of about
140 parsecs \citep*[see, e.g.,][]{reipurth3ea97}.  It consists of a
visible stellar component, T Tau N, and an infrared source T Tau S
\citep[discovered by][]{dyck3ea82}, located about 0\farcs62 south of T
Tau N.  \citet{robbertoea95} presented coronagraphic mages of the
complex emitting environment around T Tau.  Spectroscopically, SB99
identified two separate outflows moving almost perpendicular to each
other, consisting of an E--W outflow \citep[e.g., HH 155, discovered
by][]{schwartz75} originating from T Tau N and a N--S outflow
originating from the embedded source, T Tau S.  \citet{reipurth3ea97}
proposed that the N--S outflow is the likely source for the giant
outflow HH 355.

In their high spatial and spectral resolution spectroscopic study of
the T Tau environment, BS94 identified several kinematic components
(A, B, C, D, and E) of the outflows.  Component E coincides with HH
255 and lies south of T Tau, along the N--S outflow.  The components
of the N-S outflow near T Tau are apparent in Figure
\ref{fig_bohmsolf94}, a position-velocity diagram from the [S II] 6731
line at a position angle of $0^\circ$/$180^\circ$.  SB99 carried out a
deeper spectroscopic study and identified two more components, I and
J, of the N--S outflow.  Component I lies about 10\arcsec\ north and J
lies about 15\arcsec\ south of T Tau.  By considering only the
components C, D, I, and J, SB99 derived an inclination of $\sim
11^\circ$ from the plane of the sky for the N--S outflow.  This
inclination implies a total outflow velocity of $\sim 90$ km s$^{-1}$
for I and J and $\sim 280$ km s$^{-1}$ for C and D.

HH 255 (component E) lies along the outflow south of T Tau, but it
possesses unusual kinematic behavior (as noted in \S
\ref{sec_introduction} and described below).  There are two
possibilities: Either HH 255 is physically associated with the T Tau
outflow, or it is simply projected onto the region between
condensations D and J and coincidently has the same radial velocity as
T Tau.  In the unlikely event that it is a projected object, HH 255
remains mysterious (i.e., it still exhibits an HH spectrum with
unusual kinematic behavior).  In addition, the position-velocity
diagram (Fig.\ \ref{fig_bohmsolf94}) contains possible evidence for a
physical connection with component D.  First, HH 255 seems to begin at
the same place where D disappears (D extends to about 4\arcsec, while
E spans from roughly 4\arcsec\ to 10\arcsec\ south of T Tau).  Also,
BS94 note an apparent ``bridge'' of emission between D and E in the
position-velocity diagram (most evident in [N II]), suggesting a rapid
decline in radial velocity at the edge of component D that
``connects'' to the low radial velocity of HH 255.  Therefore, we will
hereafter assume that HH 255 is physically associated with the outflow
south of T Tau.

The spectrum of HH 255 is consistent with a moderately high excitation
HH object \citep*[SB88;][]{raga3ea96}, suggesting that it is emission
from shock-excited gas.  Further evidence for a shock was found by
BS97, who derived information about the ionization, centroid velocity,
and velocity dispersion from emission lines in components D and E.
They found an essentially discontinuous rise (as one moves south) of
the line ratios [O III]/H$\alpha$, [N II]/[N I], and [O II]/[O I]
between a distance of 3\arcsec\ and 4\arcsec\ south of T Tau,
indicating a drastic increase in ionization there.  Beyond that, the
ionization is roughly constant from 4\farcs5 to 8\farcs5.  The
centroid velocity (of, e.g., [N II]) has maximum (blueshift) of
roughly $-45$ km s$^{-1}$ at 2\arcsec\ (inside component D), decreases
to $-25$ km s$^{-1}$ from 2\farcs5 to 3\farcs5, and then plummets to
basically zero from 3\farcs5 to 4\farcs5.  The centroid velocity
remains near zero (ranging from $0$ to $-5$ km s$^{-1}$) from 4\farcs5
to about 10\arcsec\ (within HH 255).  The decreasing centroid velocity
and increasing ionization (as one moves from D to E) suggests the
existence of a shock that faces toward T Tau.  That is, material is
rapidly decelerated at a position of $\sim 4\arcsec$, and the kinetic
energy of the flowing material is converted into thermal energy via a
shock.

The presence of [O III] in the spectrum of HH 255 requires a shock
velocity (i.e., pre-shock velocity minus post-shock velocity) of $>90$
km s$^{-1}$ \citep{hartigan3ea87}.  The difference in the deprojected
velocities of component D and HH 255 allow for such a shock velocity.
However, the shock model becomes complicated when one considers the
behavior of the velocity dispersion in the shocked region.  BS97 found
that velocity dispersion (the full width at half maximum of [N II] and
[S II]) falls from about 80 km s$^{-1}$ at 3\farcs5 to about 25 km
s$^{-1}$ at 4\farcs5 and remains the same to about 9\arcsec\ south of
T Tau.  Bow shock models \citep[e.g.,][]{hartigan3ea87, ragabohm85,
ragabohm86} predict that the velocity dispersion should increase where
the centroid velocity decreases.  In contrast, at the interface
between component D and E, both the centroid velocity and the velocity
dispersion decrease at the same location.  Further, the bow shock
models predict that the full width at zero intensity (FWZI) of
observed lines should be equal to the shock velocity
\citep{hartigan3ea87, ragabohm85, ragabohm86}.  In HH 255, the FWZI
(as estimated by the full width at 10\% intensity) remains constant at
about 45 km s$^{-1}$ (BS97), well below the minimum shock velocity of
$90$ km s$^{-1}$ required by the [O III] flux.

This means that HH 255 is not a bow shock.  Unlike typical HH objects
\citep{hartiganea00, raga3ea96}, it is not a working surface within or
at the head of a jet or bullet, and it cannot be a shock around a
stationary cloudlet.  On the other hand, the fact that the centroid
velocity is almost zero and the velocity dispersion is very small
suggests that HH 255 is a standing shock with some shape other than a
bow.  We will discuss this possibility further in section
\ref{sec_discussion}.  It is clear that, in order to understand HH
255, we have to study additional observational properties of this
emission region.  So far, most of its enigmatic properties have been
deduced from the kinematics and ionization, so in the next section, we
use existing data to derive information about the line flux and
density within HH 255.

\section{Line Flux and Density Stratification  \label{sec_further}}

In order to gain more insight, we will look at the emission line
fluxes as a function of position along HH 255 derived from previous
observations.  We focus specifically on the well-observed lines
H$\alpha$, [O I] 6300, [S II] 6731, [N II] 6583, and [O III] 5007.  We
know that components D and J are parts of the southern outflow, which
moves nearly perpendicular to the line of sight (SB99), and we will
assume that HH 255 (component E) is a part of that outflow.  We shall
also make use of the fact that HH 1 is a well-studied and relatively
well-behaved bow shock \citep[SBR88;][]{choe3ea85, hartigan3ea87,
solfbohm91, reipurthbally01, ballyea02}.  Since HH 255 and HH 1 have
the same overall excitation, and both are seen nearly side-on, one
would naively expect that the line flux stratification of each line
would be similar for both objects.  Therefore, in the following study,
we will compare HH 255 to HH 1.  One must be aware that typical HH
objects conform to a simple bow shock model only very approximately,
at best, so we do not expect more than a crude, qualitative agreement
between the two objects.  Also, the enigmatic ionization structure and
kinematic properties of HH 255 (discussed in \S \ref{sec_properties})
already show that it is unique.

The observational data that we need for this investigation can be
extracted from the earlier studies of SBR88, BS94, BS97, and SB99.
For instance, SBR88 present information about the ratios [O
I]/H$\alpha$, [S II]/H$\alpha$, [N II]/H$\alpha$, and [O
III]/H$\alpha$, and about H$\alpha$ itself, along a line from 2\farcs7
to 10\farcs0 south of T Tau.  Additional information about some of the
lines (with higher spatial and spectral resolution) has been extracted
from BS94.  Information about line emission at larger distance from T
Tau (which we used only indirectly) is available in SB99.

\begin{figure}
%\plotone{figs/HH255_2lines.eps}
\plotone{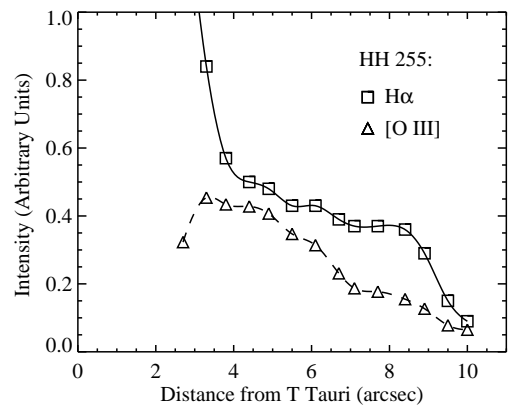}

\caption{Line fluxes of H$\alpha$ and [O III] 5007 in HH 255 as a
function of position south of T Tauri.
\label{fig_HH255_2lines}}

\end{figure}

In Figure \ref{fig_HH255_2lines}, we show the H$\alpha$ and [O III]
line fluxes as a function of position south of T Tau.  We focus on the
region containing the transition between component D and HH 255 (from
$\sim 3\arcsec$--4\farcs5) and also following the entire extent of HH
255.  The fluxes of each line are given in arbitrary units (note that
line ratios derived from this plot would not be correct).  It is
evident in Figure \ref{fig_HH255_2lines} that [O III] peaks in the
transition region and decreases (though gradually) with distance
inside HH 255.  Component D (which is an important part of the N--S
outflow; see \S \ref{sec_properties}) does not contribute
significantly to the [O III] emission.  On the other hand, the
H$\alpha$ line is relatively bright in component D, which is fairly
close to T Tau, and decreases rapidly to the northern edge of HH 255.
Within HH 255, H$\alpha$ also decreases gradually with distance from T
Tau before decreasing rapidly again at the southern edge of HH 255.

\begin{figure}
%\plotone{figs/HH1_2lines.eps}
\plotone{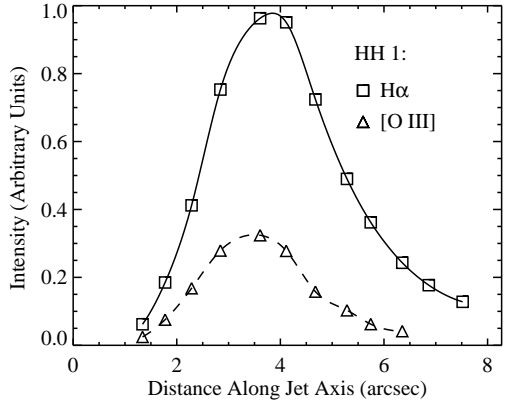}

\caption{Line fluxes of H$\alpha$ and [O III] 5007 in HH 1 as a
function of position along the extrapolated jet axis.  The distance
increases toward the location of the jet source.
\label{fig_HH1_2lines}}

\end{figure}

For comparison, we plot the H$\alpha$ and [O III] line fluxes in HH 1
along the extrapolated jet axis in Figure \ref{fig_HH1_2lines}.  Since
the HH 1 bow shock points away from the outflow source (in contrast to
HH 255, see \S \ref{sec_properties}), we have plotted the distance in
arcseconds, from an arbitrary starting point, moving toward the source
(i.e., in the opposite sense as in Fig.\ \ref{fig_HH255_2lines}).  We
have chosen this orientation so that pre-shock material is plotted on
the left hand side of both Figures \ref{fig_HH255_2lines} and
\ref{fig_HH1_2lines}, and post-shock material is on the right.  There
are two things to consider when comparing the apparent length scales
in these two Figures.  First, HH 1 is about three times further away
than T Tauri.  Second, the characteristic length scales (e.g., the
cooling length) of each object will be different because of the
different pre-shock particle densities and the different sizes and
shapes of the jets that drive each shock.  The two effects partially
cancel to give an apparent spatial size of HH 1 that is roughly
similar to HH 255.

A comparison between Figures \ref{fig_HH255_2lines} and
\ref{fig_HH1_2lines} reveals a qualitative agreement in [O III].  Both
objects exhibit a rise in [O III] and a subsequent decline.  The
decline of [O III] in HH 255, however, is apparently slower than in HH
1.  The H$\alpha$ emission in HH 1 behaves the same as [O III].  This
is not the case for HH 255.  The steep decline of H$\alpha$ to the
northern edge of HH 255 is typical of many T Tauri stars and probably
has nothing to do with the transition region between component D and
HH 255.  The decline of H$\alpha$ within HH 255 is similar to the
decline of H$\alpha$ in HH 1.  However, the decline is apparently
slower in HH 255, and beyond 8\farcs5, the H$\alpha$ emission drops
off suddenly much more steeply.  The increase in [O III] and gradual
decline of [O III] and H$\alpha$ (present in both objects) conforms to
the general model of a cooling/recombination region behind a shock.
This rough agreement between Figures \ref{fig_HH255_2lines} and
\ref{fig_HH1_2lines} strengthens the assumption that HH 255 arises
from emission behind a shock that faces toward the outflow source,
while the HH 1 bow shock points away from its source.

\begin{figure}
%\plotone{figs/HH255_3lines.eps}
\plotone{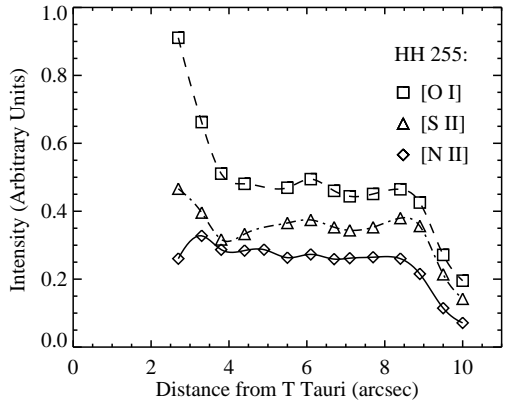}

\caption{Line fluxes of [O I] 6300, [S II] 6731, and [N II] 6583 in HH
255 as a function of position south of T Tauri.
\label{fig_HH255_3lines}}

\end{figure}

\begin{figure}
%\plotone{figs/HH1_3lines.eps}
\plotone{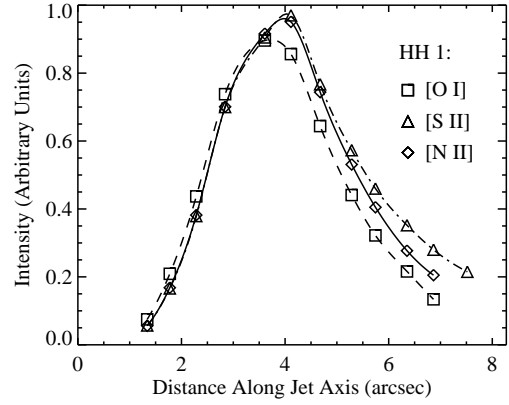}

\caption{Line fluxes of [O I] 6300, [S II] 6731, and [N II] 6583 in HH
1 as a function of position along the extrapolated jet axis.  The
distance increases toward the location of the jet source.
\label{fig_HH1_3lines}}

\end{figure}

Figures \ref{fig_HH255_3lines} and \ref{fig_HH1_3lines} contain the
flux distribution for HH 255 and HH 1 (respectively) of the [O I], [S
II], and [N II] lines. In these lines, the fluxes along HH 255 (Fig.\
\ref{fig_HH255_3lines}) are drastically, though systematically,
different from the results in HH 1 (Fig.\ \ref{fig_HH1_3lines}).  The
behavior of [O I], [S II], and [N II] in HH 1 (Fig.\
\ref{fig_HH1_3lines}) is basically the same as for [O III] and
H$\alpha$ (discussed above; Fig.\ \ref{fig_HH1_2lines}) and represents
the expected behavior for a bow shock.  By contrast, the emission
lines in Figure \ref{fig_HH255_3lines} remain surprisingly constant
(to within 10\% for [N II], 18\% for [S II], and 9\% for [O I]) along
the extent of HH 255, between 4\arcsec\ and 9\arcsec\ south of T Tau
(corresponding to a distance of $\sim 700$ AU).  Beyond this point,
the fluxes decrease steeply.  The steep decrease occurs in the same
place for all three lines (and for H$\alpha$ in Fig.\
\ref{fig_HH255_2lines}), so it cannot be a consequence of a change in
ionization.  This could be the consequence of a drastic increase in
the line of sight extinction at that location, but the most likely
explanation is a rapid decline in the general particle density at
about 9\arcsec\ south of T Tau.  As far as we know, this behavior has
not been seen before in any part of an outflow from a T Tauri star.

\begin{figure}
%\plotone{figs/HH255_Ne.eps}
\plotone{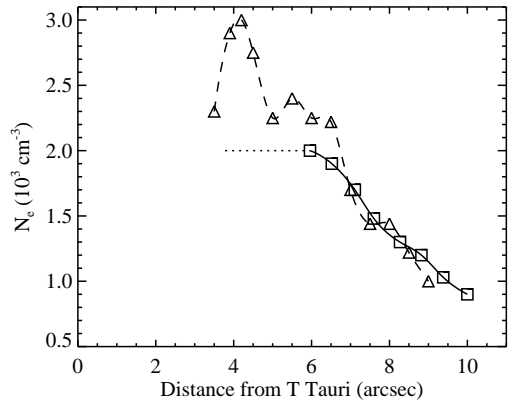}

\caption{Electron density in HH 255 as a function of position south of
T Tauri, deduced from the [S II] 6731/6716 line ratio.  The squares
are from \citet{solf3ea88}, and the triangles are from
\citet{bohmsolf94}.
\label{fig_HH255_Ne}}

\end{figure}

The unusual fact that the emission line fluxes of [O I], [S II], and
[N II] remain constant throughout HH 255 should be discussed further.
Remember that these forbidden emission lines are generated by
radiative de-excitation to the ground state by atoms that have been
excited by collisions with free electrons \citep[see,
e.g.,][]{osterbrock89}.  In this context, the electron density,
$N_{\rm e}$, is an important quantity, so we have plotted $N_{\rm e}$
as a function of position south of T Tau in Figure \ref{fig_HH255_Ne}.
The two lines in the Figure represent data from two studies (SBR88;
BS94), and they are in agreement with each other (though the
measurements were taken at different epochs).  Note that the electron
density in HH 255 is relatively low (always $< 3\times 10^{3}$
cm$^{-3}$), so collisional de-excitation is not an important process.
Since the ionization state of the gas is roughly constant within HH
255 (BS97), a simple explanation for the constancy of emission lines
would be that $N_{\rm e}$ and the particle density are also constant
throughout.  It is clear from Figure \ref{fig_HH255_Ne}, however, that
this explanation is not acceptable because $N_{\rm e}$ decreases by a
factor of two or three along HH 255.

The volume emissivity (emission per cm$^{3}$) for a given line of a
given species is proportional to $N_{\rm e} N_{\rm s}$, where $N_{\rm
s}$ is the number density of the species.  The measured fluxes shown
in Figure \ref{fig_HH255_3lines} represent the emission per cm$^{2}$
(since all of the emission adds along the line of sight).  The fact
that $N_{\rm e}$ decreases, while the line fluxes are constant,
requires a strict relationship between the geometry, velocity, and/or
time-dependence of the outflow within HH 255.  In the next section, we
will discuss a plausible model for the flow that explains some (but
not all) of the observations.

%For example, in a steady-state flow (where the mass is conserved as a
%funciton of position in the flow), the geometry and velocity of the
%flow must follow a strict relationship, in order for the line flux to
%be constant at the same place where $N_{\rm e}$ decreases.  Assuming a
%constant temperature, there are only a few possible explanations.  If
%$N_{\rm e}$ decreases, the volume emissivity must also decrease,
%unless $N_{\rm s}$ increases at the same time.  This can occur in a
%region where the ionization is decreasing while, at the same time, the
%overall density is increasing.  This could conceivably happen for a
%flow that is converging, decelerating, or both.  If the flow is
%divergent, however, $N_{\rm e}$ and $N_{\rm s}$ decrease together
%(ignoring recombination), and the volume emissivity will decrease as
%$N^2$ (or steeper, with recombination).  In this case, the measured
%line flux will be constant only if the flow is decelerating in a
%manner such that the decreasing volume emissivity is exactly canceled
%by an increasing column density.  The details of the flow are altered
%if we consider changes in temperature, but the required relationship
%between the geometry and velocity remains strict.

%If the flow is time-dependent, there are more possible explanations.
%Note that the data indicates a sudden drop in the matter density at
%the southern end of HH 255 (as discussed above).  Unless there is a
%drastic divergence or acceleration of the flow at this location (which
%seems unlikely), the flow there must be time-dependent.  

\section{Discussion \label{sec_discussion}}

\subsection{Summary of Properties \label{sub_summary}}

We have presented convincing evidence that the northern edge of HH 255
is a reverse-facing shock in the southern outflow of T Tau.  The
kinematic structure of the shock (see \S \ref{sec_properties}) is
drastically different than expected for a bow shock, which is the
successful model for typical HH objects.  The emission region behind
the shock, which comprises the extent of HH 255 itself, is also
unusual, requiring a flow structure in which the electron density
decreases, while some of the line fluxes remain constant over a large
fraction of the flow (see \S \ref{sec_further}).  Some of the basic
properties of HH 255 can be summarized as follows:
\begin{enumerate}

\item{The ionization state is consistent with emission excited by a
shock, and the strength of [O III] 5007 emission requires a shock
velocity of greater than 90 km s$^{-1}$.  The ionization is roughly
constant along the extent of HH 255.}

\item{The radial velocity, $v$, goes nearly to zero across the shock,
at the same place where the velocity dispersion, $\Delta v$,
decreases.  Behind the shock, $\Delta v$ remains less than 45 km
s$^{-1}$ along the extent of HH 255.  In contrast, bow shock models
predict that the $\Delta v$ increases where the $v$ decreases, and
that $\Delta v$ is equal to the shock velocity.}

\item{The electron density, $N_{\rm e}$, has a maximum at the location
of the shock, then decreases by a factor of 2--3 along the extent of
HH 255 (see Fig.\ \ref{fig_HH255_Ne}).}

\item{The fluxes of the [O I] 6300, [S II] 6731, and [N II] 6583 lines
are constant along the extent of HH 255.}

\item{At the southern edge of HH 255, the fluxes of the [O I], [S II],
[N II], and H$\alpha$ lines decrease drastically and simultaneously.}

\end{enumerate}

\subsection{A Preliminary Model \label{sub_model}}

We would like to find a model that, at least partially, explains the
enigmatic properties of HH 255 listed above.  To that end, we propose
that HH 255 results from a standing or quasi-stationary shock,
possibly associated with the collimation of the southern outflow (or
part of that flow) from T Tau.  A basic introduction of standing
shocks, and the distinction between a receding and a stationary shock,
was originally given by \citet{courantfriedrichs48}.

There is a theoretical justification for expecting reverse-facing,
slow-moving shocks near the collimation region of outflows from YSO's.
Models for the launching of such outflows \citep[for a review,
see][]{shibatakudoh99, koniglpudritz00} rely on magnetic processes
within the wind that lead to a self-collimation of material near the
axis into a jet.  However, not all of the material in the launched
wind is collimated, so the models predict outflows with both a
collimated and a wide-angle component \citep[e.g., ][]{lishu96}.
Observational support for a two component wind very near the central
star (i.e., a jet plus a wide-angle ``disk wind'') was noted by
\citet{kwantademaru88}.  Recently, a few authors
\citep*{franknoriegacrespo94, frankmellema96, mellemafrank97,
delamarter3ea00, gardiner3ea02, matt02, matt3ea03} have shown that the
interaction between the wide-angle wind and the circumstellar
environment results in a reverse-facing shock that is stationary or
moves slowly with respect to the head of the collimated jet
\citep*[see also][]{frank3ea02}.  The reverse shock completely
encloses the wind and bears no morphological or kinematic resemblance
to a bow shock.  These authors have also shown that, depending on the
details of the wind and its environment, the wide-angle wind can
itself become collimated by thermal or magnetic pressure gradients in
the post-shock region, adding to the total collimated flow and leading
to broader and more powerful jets than in the pre-shock, free-flowing
wind.

It is not clear what the observational signature of this reverse-shock
would be. To our knowledge, no one has claimed to observe such a
feature in a YSO outflow \citep*[though such a possibility is
discussed in the very recent work of][]{bally3ea02}.  Presumably, the
reverse shock exists relatively close to the source (within a distance
that is comparable to the width of the collimated outflow) and would
therefore be difficult to disentangle from the complex emitting
environment there.

To get more insight into this problem, and to serve as an illustrative
example, we have run a very simplified numerical magnetohydrodynamic
(MHD) simulation of the interaction between an isotropic central wind
and a constant, vertical magnetic field held fixed on a plane (as if
fixed in a disk).  This work is similar to that of \citet{matt02}, and
we have used their 2.5D (axisymmetric) MHD simulation code \citep[the
code is also described in][]{mattea02}.  The code uses a two-step
Lax-Wendroff (finite-difference) scheme to solve the equations on a
group of nested, Eulerian grids.  For this very simple, illustrative
case, we consider only the effects of ideal MHD on an adiabatic
plasma, and we ignore the effects of gravity and rotation.

For the central wind boundary condition, we use a sphere with a radius
of 40 grid points, corresponding to 40 AU.  There, the wind turns on
at $t = 0$ and is held constant and radial with a velocity of 280 km
s$^{-1}$ (with a Mach number of 5) and a mass outflow rate of $2
\times 10^{-8} M_\odot$ yr$^{-1}$ (consistent with the known
properties of the southern T Tau outflow and with typical HH outflows;
\citealp{reipurthbally01}).  The rest of the grid is initialized with
a constant density of 60 particles per cm$^{3}$ and a constant
magnetic field of $10^{-3}$ Gauss (the ratio of the Alfv\'en to sound
speed in the ambient environment is 5).  At the cylindrical $z = 0$
boundary, we held the magnetic flux constant and did not allow outflow
or inflow onto the boundary.  In this way, the magnetic field lines
are ``anchored'' at the equator as if embedded in a geometrically
thin, conducting disk.  We use outflow boundary conditions on the
outer boundaries of our outermost grid (of which there were 5, nested
concentrically).

When the simulation begins, the wind propagates outward, and sweeps up
the ambient material and vertical magnetic field ahead of it.  With
the chosen parameters, the kinetic energy in the wind (which goes
roughly as $r^{-2}$) becomes equal to the ambient magnetic energy
density (which is constant for a vertical field) at a radius of about
500 AU.  At this location, there is a reverse-facing shock where the
wind converts some of its kinetic energy to thermal energy.  This
shock is stagnant, and the steady-state solution contains a jet
composed of post-shock wind gas.  Figure \ref{fig_exaflow} shows the
density (logarithmic), velocity vectors, and magnetic field lines of
this steady-state solution achieved in the simulation.

The Figure is not intended to represent the southern outflow of T Tau
in any detail, but it illustrates some of the key features of our
proposed explanation.  Most importantly, there is a reverse-facing
shock that is stagnant at a position of about 500 AU (corresponding to
3\arcsec\--4\arcsec\ at the distance to T Tau) along the $z$ direction
in Figure \ref{fig_exaflow}.  The shock does not resemble a bow shock
morphologically nor kinematically, as post-shock material does not
have a large velocity dispersion as it moves essentially vertically
within the jet.  Since the magnetic field is fixed on the equator (in
this particular example), the flow at wide angles hits an oblique
shock, which compresses and forms a sort of shell of relatively fast
moving material around the flow that continues past 500 AU.  This
feature is dependent on the details of this type of collimator, and we
do not mean to suggest that such a feature exists in HH 255.  In the
next section, we will compare the expected behavior of the proposed
stationary shock to the observed properties of HH 255.

\subsection{Comparison of Model to HH 255}

\begin{figure}
%\plotone{figs/old_bohmsolf94.eps}
%\plotone{figs/new_bohmsolf94.eps}
\plotone{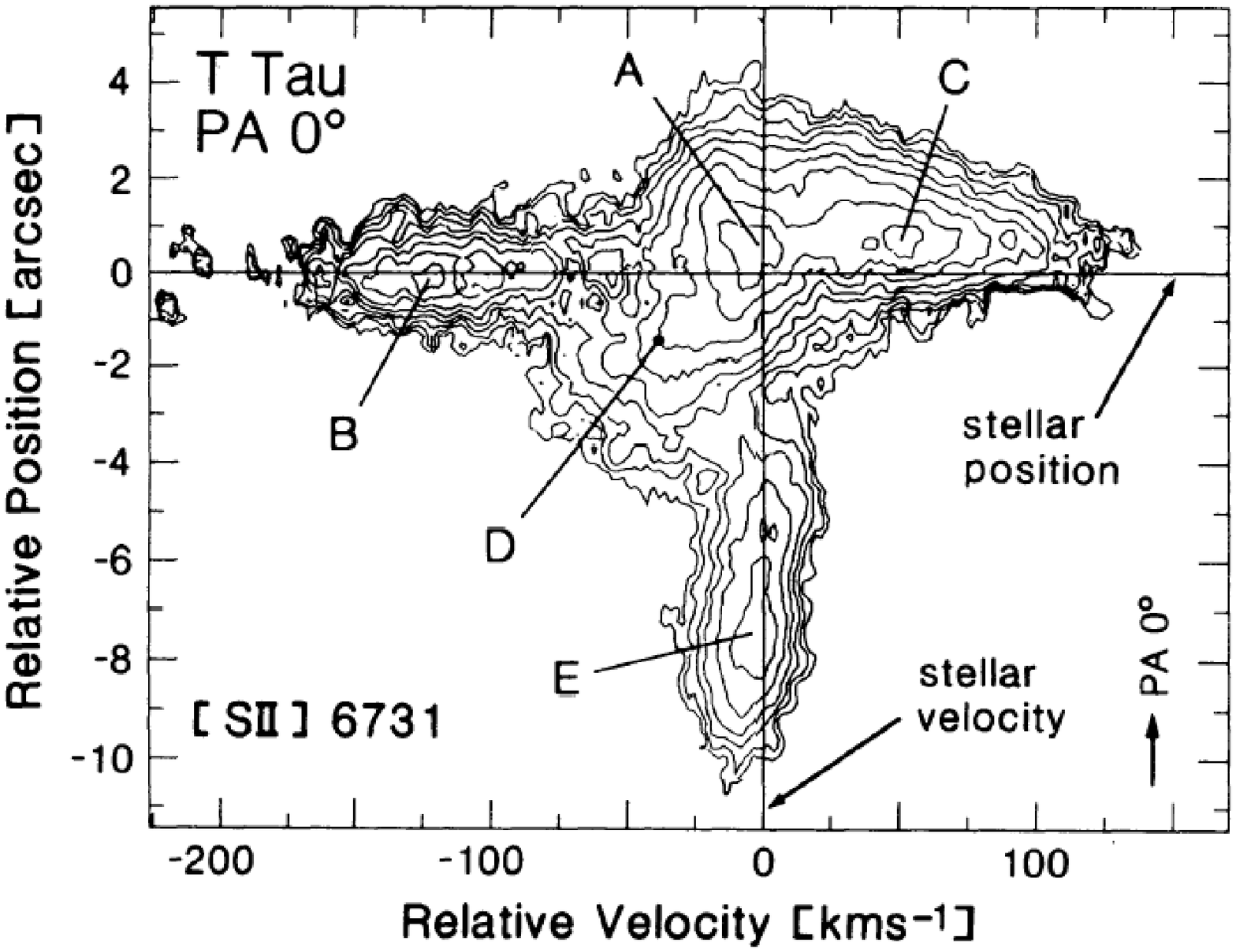}

\caption{Position-velocity diagram extracted from an echelle
observation of the [S II] 6731 line along the position angle of
$0^\circ$ and centered on T Tauri.  This figure was adapted from
figures 1b and 2b of \citet{bohmsolf94}, who identified the
spatio-kinematic components A, B, C, D, and E.
\label{fig_bohmsolf94}}

\end{figure}

Figure \ref{fig_bohmsolf94} shows that, at $\sim 4$\arcsec\ south of T
Tau, there is an abrupt transition in velocity.  It starts from a
centroid radial velocity of $-45$ km s$^{-1}$ in component D (which
corresponds to a total velocity of $-280$ km s$^{-1}$; see \S
\ref{sec_properties} and SB99), and decreases to only a few km
s$^{-1}$ in HH 255 (also see fig.\ 5 of BS97).  The velocity
dispersion decreases in the same region.  Furthermore, the ionization
of N and O increases drastically southward between 3\arcsec\ and
4\arcsec\ south of T Tau, and the electron density has a maximum in
the same region (BS97).

These properties at the northern edge of HH 255 would be rather well
explained by the existence of a standing shock.  In any sort of
ionizing shock, the velocity will decrease at the same place that the
density and ionization state increases.  However, the velocity
dispersion will only be low (as in HH 255) if post-shock gas flows
through the shocked region and does not diverge rapidly (i.e., to move
around some obstacle; as in the case of a bow shock).

\begin{figure}
%\plotone{figs/sigma1disklowrho_fig.eps}
\plotone{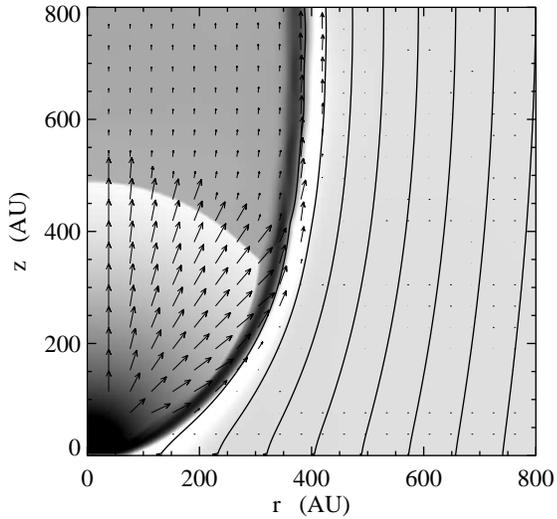}

\caption{Logarithmic density greyscale (black is highest density,
white is lowest), velocity vectors, and magnetic field lines in the
steady-state solution of the numerical simulation described in the
text.
\label{fig_exaflow}}

\end{figure}

\begin{figure}
%\plotone{figs/fig_FakePVD.eps}
\plotone{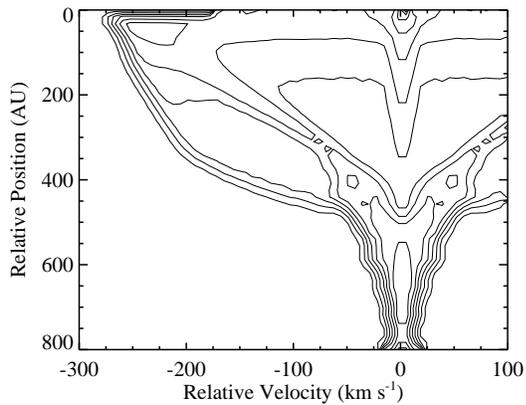}

\caption{Artificial position-velocity diagram of the data shown in
Figure \ref{fig_exaflow}.  Shown are logarithmic contours of the
velocity histogram as a function of $z$ through a 140 AU wide slit, as
seen from a line of sight perpendicular to the jet axis.  The stellar
position is at the top, and it's velocity is 0 km s$^{-1}$.
\label{fig_fakepvd}}

\end{figure}

To illustrate this point further, and for comparison to Figure
\ref{fig_bohmsolf94}, we generated an artificial position-velocity
diagram (Fig.\ \ref{fig_fakepvd}) from the simulation data in our
illustrative example discussed in section \ref{sub_model}.  To do
this, we used our steady-state solution plotted in Figure
\ref{fig_exaflow}.  With the assumption of axisymmetry, we generated a
3D data cube, placed the observer along the cylindrical $r$ direction
(i.e., the observer sees the jet in the plane of the sky), and made a
histogram of velocities as a function of $z$.  Because we are
interested in where the most emission is likely to occur, we weighted
the velocities in the histogram by the density squared.  Furthermore,
we only considered the material within a slit aperture with a width of
140 AU (corresponding to the 1\arcsec\ slit width used by BS94).  The
data has been smoothed to a spatial resolution of roughly 30 AU and we
used a velocity binning corresponding to a resolution of about 15 km
s$^{-1}$.  Figure \ref{fig_fakepvd} contains the resulting
position-velocity diagram.  Note that the vertical distance scale is
reversed, with respect to Figure \ref{fig_exaflow}.

It is clear from Figure \ref{fig_fakepvd} that, at the location of the
reverse shock ($\sim 500$ AU), the velocity dispersion becomes quite
small compared to the shock velocity of nearly 280 km s$^{-1}$.
Figures \ref{fig_bohmsolf94} and \ref{fig_fakepvd} do not agree at all
in the region closer to the star than 500 AU.  Note that, in Figure
\ref{fig_bohmsolf94}, the region near the star is dominated by the
emission from component D and emission from other parts of both
outflows (N--S and E--W) from T Tau.  In Figure \ref{fig_fakepvd}, the
structure in region near the star depends on the details of the
specific example we chose, and the very broad width (e.g., compared to
component D in Fig.\ \ref{fig_bohmsolf94}) is due to the fact that the
simulated central wind is completely isotropic (a partially collimated
central wind could have a much narrower $\Delta v$).  In spite of
this, the region from 500 AU and beyond in Figure \ref{fig_fakepvd} is
qualitatively quite similar to component E (HH 255) in Figure
\ref{fig_bohmsolf94}.  This general similarity demonstrates that a
reverse-facing, stationary shock at the position of 500 AU may explain
some of the unusual spatio-kinematic properties of HH 255.

There are other aspects of the observations, however, that are more
difficult to explain, without considering the radiative properties of
the shock.  Specifically, the details of the emission distributions of
the H$\alpha$, [O III], [N II], [S II], and [O I] emission lines
throughout HH 255 (discussed in \S \ref{sec_further}) have not been
explained.  It is not obvious that a standing shock, as discussed in
section \ref{sub_model}, should lead to a fundamentally different line
emission distribution than a bow shock.  In fact, we did find in
section \ref{sec_further} that, in HH 255, the [O III] and H$\alpha$
emission agrees somewhat with the prediction of simplified bow shocks
\citep[see also Fig.\ 9 of][]{ragabohm85}.  We have also not explained
the formation of molecular H$_2$ lines, which presumably form in low
velocity (tens of km s$^{-1}$) shocks.  However, the existence of a
reverse shock does explain how the optical forbidden emission lines
can have a $\Delta v$ that is comparable to the typical (for HH
objects) $\Delta v$ of H$_2$ lines, as is the case for HH 255 (SB99).

Perhaps the biggest problem, however, is that the fluxes of the [N
II], [S II] and [O I] lines are essentially constant, while the
electron density decreases, over a large distance within HH 255 (\S
\ref{sec_further}).  In the adiabatic simulation shown in Figure
\ref{fig_exaflow}, the density goes from 40 to 130 cm$^{-3}$ (and
velocity from 290 to 90 km s$^{-1}$) across the shock on axis.  The
density (and jet width and velocity) remains roughly constant beyond
that, so that for a constant temperature, the line fluxes in that
region would be constant.  But in a true, astrophysical shock at those
densities, radiative cooling (and ionization) would play a significant
role, leading both to a larger density jump across the shock and to a
temperature gradient behind the shock.  For example,
\citet{shullmckee79} showed that, for a pre-shock density of 10
cm$^{-3}$ and a shock velocity of 100 km s$^{-1}$, the maximum density
behind a radiating shock is 6450 cm$^{-3}$, leading to a cooling
region that is quite small (of order 10 AU).  However, if one ignores
the cooling just behind the shock (or if the post-shock density is
significantly lower), the cooling time for gas at a density of 10$^3$
cm$^{-3}$ is of the order of 100 years \citep{aller84,
shapiromoore76}, which is comparable to the crossing time of material
in HH 255.  At present, we are unable to resolve this dilemma, without
doing a detailed and self-consistent cooling (and ionization)
calculation for a reverse-facing, standing shock.  We leave such a
calculation and detailed comparison to the emission line distribution
in HH 255 for future work.

At present, it is conceivable that HH 255 is the emission behind a
reverse-facing, standing (non bow) shock, possibly associated with the
collimation of the southern outflow from T Tau.  However, other
possibilities have not been fully explored.  One potentially important
effect is that of time-dependence within the flow that formed HH 255.
For example, the fact that the emission of the lines of [N II], [S
II], [O I], and H$\alpha$ decrease simultaneously and drastically at
the southern edge of HH 255, indicates (most likely) a fast decrease
in the matter density.  This suggests that, at some earlier time, the
flow exhibited an increase in density (or velocity, or both),
resulting in the apparent density jump at the southern edge of HH 255.
A flow that is not in a steady state (i.e., where the mass continuity
equation is not satisfied along the flow) would have spatial
distribution of line emission that is different than steady-state
models.  Also, for simplicity, we have not considered any
non-axisymmetric effects.

In some respects, HH 255 has been studied in more detail than any
other HH object.  Because they noted some peculiar properties, BS94,
BS97, and SB99 have studied several physical aspects of HH 255 at high
spatial and spectral resolution.  It is quite possible that the
peculiar behavior of HH 255 is exhibited by other outflows that have
escaped being seen (or recognized) in all studies to date.  We have
shown that HH 255 may have something to do with the interaction of a
wide-angle flow with its environment, something predicted (though
often indirectly) by essentially all wind launching models.
Understanding this interaction may be crucial to our understanding of
the collimation of physically broad jets.  It would therefore be of
great significance if other objects similar to HH 255 are found.

\acknowledgments

This research was supported in part by NSF grant AST-9729096, and by
NSERC, McMaster University, and CITA through a CITA National
Fellowship.

% ---------- BIBLIOGRAPHY -------------

%\bibliography{../../references}
%\bibliographystyle{apj}

%Comment the next line out if you want figures.
%\end{document}

\end{document}